\documentclass[aps,pra,showpacs,twoside,twocolumn,10pt]{revtex4-1}
\usepackage[colorlinks=true, citecolor=red, urlcolor=blue ]{hyperref}
\usepackage{epsfig,newlfont,amssymb,amsfonts,amsmath,bm,subfigure,palatino,mathtools,amsthm,braket,times,soul}
\usepackage[normalem]{ulem}

\begin{document}

\title{Emergence of entanglement with temperature and time in factorization-surface states}

\author{Titas Chanda$^{1,2}$, Tamoghna Das$^{1,2}$, Debasis Sadhukhan$^{1,2}$, Amit Kumar Pal$^{1,2,3}$, Aditi Sen(De)$^{1,2}$, Ujjwal Sen$^{1,2}$}
\affiliation{$^1$Harish-Chandra Research Institute, Chhatnag Road, Jhunsi, Allahabad 211019, India \\
$^2$Homi Bhabha National Institute, Training School Complex, Anushaktinagar, Mumbai  400094, India\\
$^3$Department of Physics, Swansea University, Singleton Park, Swansea SA2 8PP, United Kingdom}

\begin{abstract}
There exist zero-temperature states in quantum many-body systems that are fully factorized, thereby possessing vanishing entanglement, and hence being of no use as resource in quantum information processing tasks. Such states can become useful for quantum protocols when the temperature of the system is increased,  and when the system is allowed to evolve under either the influence of an external environment, or a closed unitary evolution driven by its own Hamiltonian due to a sudden change in the system parameters. Using the one-dimensional anisotropic XY model in a uniform and an alternating transverse magnetic field, we show that entanglement of the thermal states, corresponding to the factorization points in the space of the system parameters, revives once or twice with increasing temperature. We also study the closed unitary evolution of the quantum spin chain driven out of equilibrium when the external magnetic fields are turned off, and show that considerable entanglement is generated during the dynamics, when the initial state has vanishing entanglement. Interestingly, we find that creation of entanglement for a pair of spins is possible when the system is made open to an external heat bath, interacting through that spin-pair having  a repetitive quantum interaction.
\end{abstract}

\maketitle

\section{Introduction}
\label{sec:intro}
 
Quantum phase transitions -- the qualitative change of the zero-temperature state driven by the  system parameters -- of interacting quantum spin models is one of the most striking quantum mechanical features, which cannot be seen in classical spin systems \cite{sachdev}. Over the years, several physical quantities and experimental  methods have been developed for detection and classification of these  transitions \cite{qpt-rev}.  For example, in the last decade,  the trends of quantum correlation measures, in the form of entanglement \cite{horodecki-rmp}, of the zero-temperature states of a given quantum spin model are found to be an effective tool  for  identifying  its quantum phase transitions \cite{lewenstein-rev,amico-rmp}.  It is also observed that these quantum many-body systems often possess highly entangled quantum states, which can be used to implement  quantum information processing tasks like quantum circuits \cite{one-way-qc}, quantum state transmission \cite{comun-spin}. Moreover, a number of  available solid state materials \cite{solid_xy}, along with cold-atomic substrates \cite{optical-lattice, coldatom,xxz-exp, lab_xy_ion},  nuclear magnetic resonance~\cite{nmr} and superconducting qubits~\cite{sup-qub} mimic these quantum spin models. Consequently, it has been possible to engineer these models in a controlled way with currently available technology.
 
Up to now, most of the studies in the direction of characterizing the quantum many-body systems using entanglement are restricted to analyze  either \textbf{(i)} the entanglement of the zero-temperature states to obtain the indication of quantum phases, or \textbf{(ii)} the behavior of thermal entanglement at a finite temperature, or \textbf{(iii)} the dynamics of entanglement starting with an entangled state to find out its sustainability in large time. In this paper, we investigate the thermalization and dynamics of entanglement in a quantum spin model, with  unentangled zero-temperature states as initial states. Such zero-temperature states, called the \emph{factorized states} are product states across all bipartitions, having vanishing bipartite as well as multipartite entanglement for specific values of the system parameters, also known as the \emph{factorization points} \cite{factor_old,factor_new,chanda16},  and are  considered to be unprofitable for quantum information protocols that use entanglement as resource \cite{horodecki-rmp}. Given a many-body system, it is therefore important to identify factorization points in the system parameter-space, which may also form lines,  surfaces, or volumes. At the same time,  finding a recipe for creating entanglement in these regions, and its neighboring regions, is crucial where tuning to other values of system parameters is not possible. In particular, if the zero temperature state is separable or possess a very low value of entanglement for the system parameters lying in the neighborhood of the factorization points, it is not guaranteed that the canonical equilibrium state (CES), after interacting with the global heat bath, can also have vanishing entanglement for all values of temperature (cf. \cite{arnesen,aditi_rapidcom,chanda16}). Furthermore, in the case of closed as well as open system dynamics, it is not a priori clear whether generation of entanglement in the evolved state from an initial unentangled state is possible. In this paper, we address both of these questions, and answer them affirmatively.

Paradigmatic one-dimensional quantum spin systems that encounter such product states at zero-temperature are \textbf{(i)} the anisotropic XY model with a transverse magnetic field that is uniform on all the spins (UXY model) \cite{lsm,bmpapers,qpt-books,dutta-qpt-books}, and \textbf{(ii)} the same with an additional transverse magnetic field having an alternating direction depending on the lattice sites (ATXY model) \cite{chanda16,dutta-qpt-books,alt-field-old,diep}.  Note here that the UXY model (model \textbf{(i)}) is a special case of ATXY model (model \textbf{(ii)}) and in this paper, we concentrate on both of the models, for which the thermal and time-evolved states can be  analytically obtained by successive applications of Jordan-Wigner and Fourier transformations \cite{chanda16,dutta-qpt-books}. For specific values of the anisotropy parameter  and the relative strengths of the uniform and alternating transverse magnetic fields, the ground state of this model is known to be doubly degenerate and factorizable  along two hyperbolic lines, known as the \textit{factorization lines} (FL) \cite{chanda16}.  

Starting from a zero-temperature factorized state of the ATXY model, we investigate the thermal as well as dynamical properties of entanglement under two different scenarios. \textbf{(a)} The first situation is when a CES of a given spin Hamiltonian undergoes a closed unitary evolution due to a disturbance in the system parameters that drives the system out of equilibrium. \textbf{(b)} The second case deals with a system that is exposed to an external thermal bath acting as an environment. Specifically, fixing system parameters on the factorization surface, we observe that entanglement of a thermal state undergoes a double revival and collapse over varying temperature when the value of the anisotropy parameter of the one-dimensional ATXY model is chosen in the appropriate range. We comment on how the zero-entanglement region over the phase plane of the model develops entanglement with an increase in temperature as well as under a time evolution of the  system, and demonstrate that the results are not modified if one considers a finite-sized system, achievable by current technology \cite{finite_size_exp} instead of a quantum spin-chain in the thermodynamic limit.

We show that for lower values of the relative strength of the uniform transverse field, the entanglement generated in the evolved states starting from the factorized state
may oscillate at first, and then saturate at a non-zero value at the long time limit. In contrast, for high values of the field-strength, the oscillation of entanglement dies out comparatively quicker than the former case, and entanglement vanishes as time increases. In the case of higher values of the anisotropy parameter, the initial oscillation of the generated entanglement for higher values of the uniform magnetic field sustains longer. It turns out that in closed evolution, entanglement can only be preserved for a long time when the system is close to the UXY model. We also consider the open system dynamics of the model by studying the  evolution of the system in contact with external heat-baths at a different temperature, which  interact with the system through a set of chosen spins via a repetitive quantum interaction \cite{rqim, rqim_attal}. Interestingly, the open system dynamics is found to distinguish between the spin in the system that is directly connected to the external heat-bath and the spin having no interaction with the bath. In particular, thermal and temporal entanglement generation over factorized states favors those spin-pairs in the spin-chain which is in contact with the thermal bath, having moderate temperature. Moreover, we show that in the case of open system dynamics, for all values of the uniform field, lower values of anisotropy parameters are profitable in terms of longer sustenance of the generated entanglement. The advantages of our results become prominent in a situation where one is forced to prepare a physical system  in a parameter regime that corresponds to a state having almost vanishing entanglement.

The paper is organized as follows.  A brief overview of the quantum spin model under consideration, its phase diagram, and the specifications of the factorized states at zero temperature is provided in Sec. \ref{sec:model}. The emergence of thermal entanglement in quantum states corresponding to factorization points in the parameter space of the system is discussed in Sec. \ref{sec:ent_fact_line}. Sec. \ref{sec:dyn_fact_ent} reports the dynamical properties of the thermally emergent entanglement at factorization points, under closed unitary evolution as well as open system dynamics. Finally, Sec. \ref{sec:conclud} has concluding remarks.

\section{The model}
\label{sec:model}

To investigate the thermal and dynamical behavior of entanglement emerging over factorized states, we choose a one-dimensional (1D) quantum spin model consisting of $N$ spin-$\frac{1}{2}$ particles. The Hamiltonian of the model is given by  \cite{chanda16, dutta-qpt-books, alt-field-old}
\begin{eqnarray}
\hat{H}_S(t)&=&\frac{1}{4}\sum_{i=1}^N J \left\{(1+\gamma)\hat{\sigma}_i^x\hat{\sigma}_{i+1}^x+(1-\gamma)\hat{\sigma}_i^y\hat{\sigma}_{i+1}^y\right\}\nonumber \\ &&+\frac{1}{2}\sum_{i=1}^N h_i(t)\hat{\sigma}_i^z,
\label{eq:ham_atxy}
\end{eqnarray} 
where $J$ is the strength of the exchange interaction, $\gamma (\neq 0)$ is the $x-y$ anisotropy, and $\{\hat{\sigma}_i^\alpha;\,\alpha=x,y,z\}$ are the Pauli spin matrices corresponding to the spin located at the site $i$. Here, $h_i(t) = h_1(t) + (-1)^i h_2(t)$ is the site-dependent external magnetic field, having two components,
$h_1(t)$ and $h_2(t)$, which are respectively the strength of a transverse magnetic field in the $+z$ direction, and that of a transverse magnetic field in the direction $+z$ or $-z$, depending on whether the site is even, or  odd. We consider periodic boundary condition, i.e., $\hat{\sigma}_{N+1}\equiv\hat{\sigma}_1$ throughout this paper, and choose the time-dependent magnetic field to be of the form
\begin{eqnarray}
 h_1(t)= \left\{
 \begin{array}{cc}
 h_1, & t\leq 0  \\
 0, & t>0
\end{array}\right.,\;\;
 h_2(t)= \left\{
 \begin{array}{cc}
 h_2, & t\leq 0  \\
 0, & t>0
\end{array}\right..
\label{eq:magnetic_field}
\end{eqnarray} 
The implications of the specific form of the magnetic field will be clear in subsequent discussions.

In the thermodynamic limit  ($N\rightarrow \infty$), by successively applying Jordan-Wigner and Fourier transformations \cite{chanda16}, the Hamiltonian in Eq. (\ref{eq:ham_atxy}) can be rewritten in the momentum space as $\hat{H}_S(t) =\sum_{p=1}^{N/4}\hat{H}_p (t)$, where 
\begin{eqnarray}
 \hat{H}_p (t) &=& J\cos{\phi_p}(\hat{a}_p^{\dagger}\hat{b}_p+a_{-p}^{\dagger}
 \hat{b}_{-p}+\hat{b}_p^{\dagger}\hat{a}_p+\hat{b}_{-p}^{\dagger}\hat{a}_{-p})\nonumber \\
 &&-iJ\gamma\sin{\phi_p}(\hat{a}_p^{\dagger}\hat{b}_{-p}^{\dagger}+\hat{a}_{p}b_{-p}-\hat{a}_{-p}^{\dagger}\hat{b}_p^{\dagger}-\hat{a}_{-p}a_{p})\nonumber\\
 &&+ (h_1(t)  + h_2(t))(\hat{b}_p^{\dagger}\hat{b}_p+\hat{b}_{-p}^{\dagger}\hat{b}_{-p}) \nonumber \\
&&+ (h_1(t) - h_2(t))(\hat{a}_p^{\dagger}\hat{a}_p+\hat{a}_{-p}^{\dagger}\hat{a}_{-p}) -2h_1(t), 
 \label{eq:hp}
\end{eqnarray}
with $\hat{a}_{p}^\dagger$ and $\hat{b}_{p}^\dagger$ given by
\begin{eqnarray}
 \hat{a}_{2j+1}^{\dagger}&=&\sqrt{\frac{2}{N}}\sum_{p=-N/4}^{N/4}\exp{\big(i(2j+1)\phi_p \big)}\hat{a}_p^{\dagger}, \nonumber \\
 \hat{b}_{2j}^{\dagger} &=&\sqrt{\frac{2}{N}}\sum_{p=-N/4}^{N/4} \exp{\big(i(2j)\phi_p \big)} \hat{b}_p^{\dagger}.
 \label{eq:spinless_fermion}
\end{eqnarray}
Here, $\hat{a}_{2j+1}^\dagger$ and $\hat{b}_{2j}^\dagger$ are the spinless fermionic operators corresponding to the odd and even sublattices, and $\phi_p = 2\pi p /N$. Therefore, the diagonalization of $\hat{H}_S(t)$ can be achieved by the diagonalization of $\hat{H}_p$ with a proper choice of the basis.

Diagonalization of $\hat{H}_S(t)$ allows one to compute the CES and the time-evolved state (TES) while considering the dynamics of the model in the form of a closed system. The CES of the ATXY model at time $t$, is given by  $\hat{\rho}_{eq}(t)=Z^{-1}\exp(-\beta_S \hat{H}_S(t))$, with $Z=\mbox{Tr}[\exp(-\beta_S \hat{H}_S(t))]$ being the partition function, and $\beta_S=(k_{B}T_S)^{-1}$, $T_S$ being the absolute temperature of the system, and $k_B$, the Boltzmann constant. We consider a situation where the system is brought to a canonical thermal equilibrium with a heat-bath at temperature $T_S$ before the beginning of the dynamics, which we label as $t=0$. At $t>0$, the system starts evolving due to the disturbance caused by switching off the magnetic fields, as given in Eq. (\ref{eq:magnetic_field}). The evolution is governed by the Schr\"{o}dinger equation corresponding to the Hamiltonian in Eq. (\ref{eq:ham_atxy}), providing the  TES, $\hat{\rho}(t)$, at any intermediate time $t$, given by 
\begin{eqnarray}
\hat{\rho}(t)=e^{-i\hat{H}_S(t>0)t / \hbar}\hat{\rho}_{eq}(t=0)e^{i\hat{H}_S(t>0)t/ \hbar},
\label{eq:closed_evolution}
\end{eqnarray}
which can be used to compute time-variation of different physical quantities. From  $\hat{\rho}(t)$, one can obtain any reduced TES, $\hat{\rho}_\Omega(t)$, corresponding to a subsystem, $\Omega$, of the system by tracing out the rest of the parts, denoted by $\overline{\Omega}$, so that $\hat{\rho}_\Omega(t)=\mbox{Tr}_{\overline{\Omega}}[\hat{\rho}(t)]$. Using $\hat{\rho}_{\Omega}(t)$, dynamics of relevant physical quantities corresponding to the subsystem $\Omega$ can be determined. Throughout this paper, we consider a nearest-neighbor (NN) even-odd spin-pair as the subsystem $\Omega$, and the rest of the spins in the spin-chain as $\overline{\Omega}$. Dimensional analysis suggests that for the Hamiltonian $\hat{H}_S$, time $t$ in Eq. (\ref{eq:closed_evolution})  is in the unit of $\hbar /J$, and $\beta_S$ is in the unit of $1 / J$. We therefore redefine the dimensionless quantities $\beta_S$ and $t$ as  $\beta_S \rightarrow J \beta_S$ and $t\rightarrow tJ/\hbar$ respectively, and use them throughout the paper.

\begin{figure}
\includegraphics[scale=0.34]{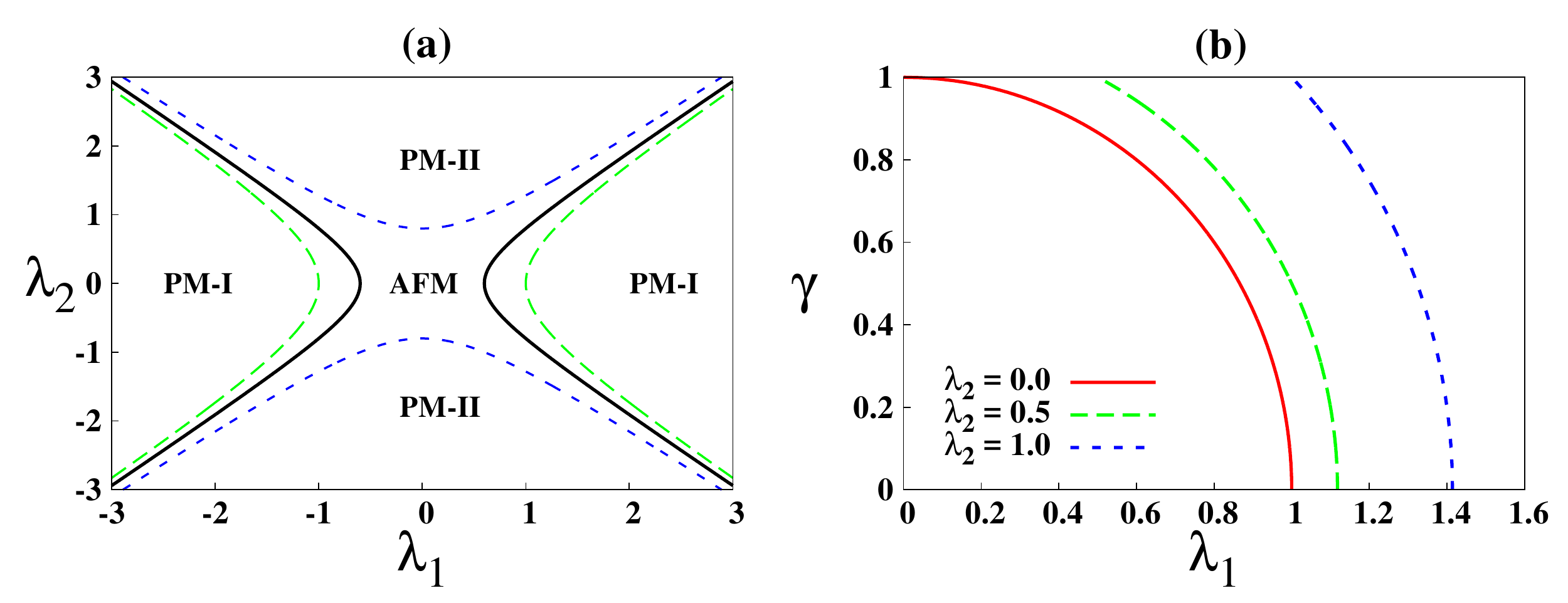}
\caption{(Color online.) \textit{Phase boundaries and factorization lines on the phase plane of the ATXY model.} (a) Phase boundaries corresponding to PM-I $\leftrightarrow$ AFM (Eq. (\ref{eq:pmafm})) and PM-II $\leftrightarrow$ AFM (Eq. (\ref{eq:dmafm})), for $\gamma=0.8$, are represented by dashed and dotted lines on the $(\lambda_1,\lambda_2)$ plane. The factorization line (Eq. (\ref{eq:fact_line})) is represented by the continuous line on the $(\lambda_1,\lambda_2)$ plane. (b)  Factorization lines corresponding to different values of $\lambda_2$ are marked on $(\lambda_1,\gamma)$ plane. The dashed and the short-dashed lines represent ATXY model, while the continuous line corresponds to the UXY model ($\lambda_2=0$). All the axes in both figures are dimensionless.}
\label{fig:phases}
\end{figure}

The ATXY model has a rich phase diagram, consisting of antiferromagnetic (AFM) and two paramagnetic (PM) (PM-I and  PM-II) phases \cite{dimertoPMII}, as depicted in Fig. \ref{fig:phases}(a)  using $\lambda_k=h_k/J$, $k=1,2$ as the system parameters in the range $\lambda_k\in [-3, 3]$ \cite{chanda16,dutta-qpt-books,alt-field-old}. In the thermodynamic limit, the boundaries between different phases in the ATXY model are given by 
\begin{eqnarray}
\lambda_1^2&=&\lambda_2^2+1 \;\;\;\;\, (\mbox{PM-I} \leftrightarrow \mbox{AFM}),
\label{eq:pmafm}
\end{eqnarray}
and
\begin{eqnarray}
\lambda_2^2&=&\lambda_1^2+\gamma^2 \;\;\; (\mbox{PM-II} \leftrightarrow \mbox{AFM}),
\label{eq:dmafm}
\end{eqnarray} 
which are also depicted in  Fig. \ref{fig:phases}(a). Note that the phase diagram is considered in a static picture at $t=0$, where the system has not started evolving in time. With $h_2(t) = 0$, Eq. (\ref{eq:ham_atxy}) reduces to the UXY model, and the PM-II phase is absent in this model. 

Apart from the phase boundaries, the variation of bipartite as well as multipartite entanglement 
suggests the existence of doubly degenerate fully separable ground states, called the \emph{factorized ground states},
in the AFM phase for specific values of $\lambda_{1,2}$ and $\gamma$. For the ATXY model, irrespective of the system-size, the factorized ground states correspond to a \emph{factorization surface} (FS), given by \cite{chanda16}
\begin{eqnarray}
\lambda_1^2=\lambda_2^2+(1-\gamma^2).
\label{eq:fact_line}
\end{eqnarray}
In Fig. \ref{fig:phases}(a), a cross-section of the FS is exhibited on the $(\lambda_1,\lambda_2)$-plane by the FLs denoted by continuous lines, in the AFM phase for $\gamma=0.8$, while in Fig. \ref{fig:phases}(b), different FLs corresponding to different values of $\lambda_2$ are depicted on the  $(\lambda_1,\gamma)$-plane. Besides indicating the phase boundaries, the NN entanglement can also efficiently indicate  the FL on the $(\lambda_1,\lambda_2)$-plane \cite{chanda16}. We will show that entanglement emerges over the FS with increasing temperature, and under time evolution in the succeeding section. 

\begin{figure*}
\includegraphics[width=\textwidth]{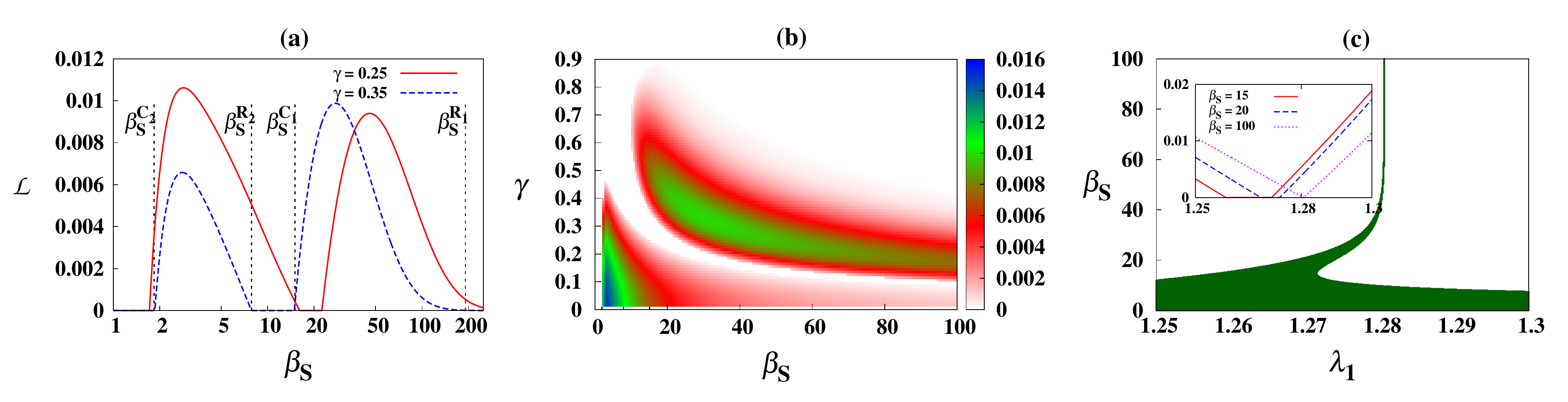}
\caption{(Color online.) \textit{Emergence of entanglement in thermal state corresponding to Hamiltonian parameters on the factorization surface.} (a) Variation of LN as a function of $\beta_S$ for different values of $\gamma$, with $\lambda_2=1$ and $\lambda_1$ being fixed by the condition of the factorization line given in Eq. (\ref{eq:fact_line}). The variation shows two successive revivals of entanglement, separated by a complete collapse, on the $\beta_S$ axes. The second revival of entanglement at $\beta_S=\beta_S^{R_2}$ is separated from the complete collapse of LN at $\beta_S=\beta_S^{C_1}$ by a finite difference, which increases with the values of $\gamma$ in the range $\gamma\leq0.45$. For $\gamma=0.35$, $\beta_S^{R_{1,2}}$ and $\beta_S^{C_{1,2}}$ are marked with vertical lines. Moreover, for $\gamma = 0.25$, we find $\mathcal{L}^{(2)}_m > \mathcal{L}^{(1)}_m$.
(b) Variation of $\mathcal{L}$ as a function of $\beta_S$ and $\gamma$, for $\lambda_2=1$, with $\lambda_1$ being fixed by Eq. (\ref{eq:fact_line}). Different shades in the figure represents different values of LN. (c) Map of the $\mathcal{L}=0$ region (shaded region) on the $(\lambda_1,\beta_S)$ plane, with $\gamma=0.6$, and $\lambda_2=1.0$. (Inset) Variation of LN as a function of $\lambda_1$ for specimen values of $\beta_S$. Note that for $\beta_S=100$, i.e., for sufficiently low temperature, the zero-entanglement region on the $\lambda_1$ axes is effectively a point, corresponding to the factorization point for fixed values of $\gamma$ and $\lambda_2$, satisfying Eq. (\ref{eq:fact_line}).  All quantities plotted are dimensionless.}
\label{fig:thermal_static}
\end{figure*}

\section{Thermal emergence of entanglement from the factorization surface}
\label{sec:ent_fact_line}

In this section, we study the static behavior of entanglement in the CES over the FS (Eq. (\ref{eq:fact_line})) in the ATXY model. Assuming the system to be a closed one, there are two extreme situations -- \textbf{(i)} the zero-temperature state (i.e., at $\beta_S=\infty$), which is fully separable on the FS, and \textbf{(ii)} the state at infinite temperature ($\beta_S=0$), which is maximally mixed, and hence with vanishing entanglement, irrespective of the values of the system parameters. For very low ($\beta_S\approx \infty$) or very high ($\beta_S\approx 0$) temperature, entanglement in the CES may still be vanishingly small due to the continuity of entanglement with the system temperature $\beta_S$. However, finding the exact region where states possess a finite amount of entanglement with increasing temperature requires careful and rigorous analysis, which will be presented here. 

Apart from these two extreme cases, thermal mixing of the entangled eigenstates of higher energy with the fully separable zero-temperature state of the Hamiltonian takes place at a moderate value of $\beta_S$. We demonstrate here that such mixing may lead to generation of entanglement  over the FS at finite system temperature. In order to do so, 
we note that the density matrix corresponding to the NN spin-pair  in  CES in the case of the ATXY model can be obtained analytically in terms of single-site magnetizations, $m^{\alpha}_{e(o)} = \mbox{Tr}(\hat{\sigma}_{e(o)}^{\alpha} \hat{\rho}_{eq}(t)),\,\alpha=x,y,z$, and two-spin correlation functions, $T^{\alpha \beta}_{eo} = \mbox{Tr} (\hat{\sigma}^{\alpha}_{e} \otimes \hat{\sigma}^{\beta}_{o} \hat{\rho}_{eq}(t)) ,\,\alpha, \beta = x, y, z$. Here, the subscripts ``$e$" and ``$o$" represent the even and odd sites respectively.  However, it can be shown that the single-site magnetizations, $m^{x}_{e(o)}$ and $m^{y}_{e(o)}$ both vanish, and the two-spin correlation functions, $T^{\alpha \beta}_{eo}=0$ for $\alpha\neq\beta$ in the case of CES.  Therefore, the two-spin density matrix corresponding to a NN spin-pair ``$eo$" corresponding to the CES is given by \cite{chanda16}
\begin{eqnarray}
\hat{\rho}_{eq}^{eo}&=&\frac{1}{4}\big[\mathbb{I}_e\otimes\mathbb{I}_o+m^z_e\hat{\sigma}^z_e\otimes\mathbb{I}_o+m^z_o\mathbb{I}_e\otimes\hat{\sigma}^z_o\nonumber\\
&&+\sum_{\alpha=x,y,z}T^{\alpha\alpha}_{eo}\hat{\sigma}^\alpha_e\otimes\hat{\sigma}^\alpha_o\big],
\end{eqnarray}
where $\mathbb{I}_{e(o)}$ is the identity matrix in the Hilbert space of the qubit ``$e$" (``$o$"). At a specific $t$, determining the values of $m^z_{e,o}$ and $T^{\alpha\alpha}_{eo}$, $\alpha=x,y,z$ at a finite system temperature $\beta_S$,  $\hat{\rho}^{eo}_{eq}$ can be computed. 


We now choose logarithmic negativity (LN) \cite{neg_group,neg_part_group} as the measure of bipartite entanglement present in an even-odd pair of NN spins. For a bipartite state $\rho_{AB}$ shared between the parties $A$ and $B$ is defined as $\mathcal{L}(\rho_{AB}) = \log_2(2 \mathcal{N} + 1)$, where the negativity, $\mathcal{N}$, is the sum of the absolute values of the negative eigenvalues of the partially transposed state, $\rho_{AB}^{T_A}$ (or $\rho_{AB}^{T_B}$), of $\rho_{AB}$ with partial transposition being taken with respect to A (or B).  We use $\hat{\rho}^{eo}_{eq}$ at $t=0$ to compute the LN in a NN even-odd spin pair as a function of the system temperature as well as the system parameters. In Fig. \ref{fig:thermal_static}, the generation of entanglement over the factorization points is demonstrated by studying the pattern of LN with respect to $\beta_S$ $(0 \leq \beta_S \leq 250)$ for different values of $\lambda_2$ and $\gamma$, where $\lambda_1$ is fixed by Eq. (\ref{eq:fact_line}). The choice of the range of $\beta_S$ is made from the observation that entanglement of the CES with $\beta_S = 250$ faithfully mimics that of the zero-temperature state. Furthermore, we observe that Fig. \ref{fig:thermal_static} reveals some interesting physics related to the theory of entanglement with the variation of the anisotropy parameter, $\gamma$, apart from establishing the primary goal of generating entanglement over the factorization points. Careful examination of Figs. \ref{fig:thermal_static}(a) and \ref{fig:thermal_static}(b) leads to the following observations.
\begin{itemize}
\item[{\bf 1.}] We first consider small values of $\gamma$, i.e., when $0 < \gamma \leq 0.45$. 

\noindent\textbf{a.} Starting from a state having vanishing entanglement at $\beta_S \gtrsim 250$, LN revives at $\beta_S^{R_1}$ and reaches a local maximum, denoted by $\mathcal{L}_{m}^{(1)}$. It then decreases and finally collapses with the increase of temperature at $\beta_S = \beta_S^{C_1}$. Interestingly, LN again revives at a higher temperature $(\beta_S = \beta_S^{R_2} < \beta_S^{R_1})$, and reaches another local maximum value,  $\mathcal{L}_{m}^{(2)}$. Finally LN collapses at $\beta_S^{C_2}$ for high values of the temperature as expected. Apart from  reestablishing non-monotonicity of entanglement with variation of system temperature, it shows a \emph{double-humped} nature of entanglement with the variation of $\beta_S$, which is rare. Note here that it is independent of the values of $\lambda_1$ and $\lambda_2$, satisfying Eq. (\ref{eq:fact_line}). Such trait of LN is depicted in Fig. \ref{fig:thermal_static}(a) for $\lambda_2 = 1$.

\noindent\textbf{b.} Moreover, we find that for certain values of $(\lambda_2, \gamma)$, $\mathcal{L}_{m}^{(2)} >  \mathcal{L}_{m}^{(1)}$ (see Fig. \ref{fig:thermal_static}(a)) even when $\beta_S$ corresponding to $\mathcal{L}_{m}^{(2)}$ is lower compared to the case of $\mathcal{L}_{m}^{(1)}$.

\item[{\bf 2.}] For higher values of $\gamma$, with the increase of the value of $\gamma$, the difference between $\beta_S^{R_2}$ and $\beta_S^{C_2}$ decreases, and eventually the \textit{double-humped} feature of the variation of LN with $\beta_S$ changes into one with a single maximum, as illustrated in Fig. \ref{fig:thermal_static}(b). Further, we observe by using numerical simulations of the Heisenberg, XXZ and XYZ models that double revivals of entanglement with temperature do not occur although single revival of the same can be obtained (see e.g. \cite{arnesen}).
\end{itemize}
 
Upto now, we have discussed how creation of NN entanglement is possible by varying temperature in the CES. Next, we study how the zero-entanglement region spreads over the phase-plane of the ATXY model with the increase in temperature. In order to investigate this, we consider the $\mathcal{L}=0$ region on the $(\lambda_1,\beta_S)$-plane with a fixed value of $\gamma$, where the value of $\lambda_2$ can be fixed, for example, at $\lambda_2=1$.   For a high value of $\beta_S$, the $\mathcal{L}=0$ region on the $(\lambda_1,\beta_S)$-plane corresponds to a specific point on the FS, which is a function of $(\lambda_1,\lambda_2)$, and $\gamma$. However, with decreasing $\beta_S$, the $\mathcal{L}=0$ point transforms into a \textit{river} on the $(\lambda_1,\beta_S)$ plane, each point in which corresponds to a thermal state of vanishing entanglement (see Fig. \ref{fig:thermal_static}(c)). The river widens and flows deeper into the AFM region with decreasing $\beta_S$  before meeting a \textit{sea} of points on the $(\lambda_1,\beta_S)$-plane corresponding to $\mathcal{L}=0$ at $\beta_S\rightarrow0$. This analysis indicates that the zero-entanglement region always remains in the AFM region on the $(\lambda_1,\lambda_2)$-plane and shifts deep inside the AFM region with the increase of temperature, making entanglement generation possible over the FL and its neighborhoods. 
The inset in Fig. \ref{fig:thermal_static}(c) shows the magnified view of the variation of LN with $\lambda_1$ for different values of $\beta_S$, when LN approaches to zero. It is evident from the figure that with decreasing $\beta_S$, the zero-entanglement region on the $\lambda_1$ axes widens, as also pointed out in the above discussion. Such a spreading of vanishing entanglement region in the AFM phase can also be illustrated by other values of $\gamma$ and $(\lambda_1, \lambda_2)$.

\section{Dynamics of emergent entanglement}
\label{sec:dyn_fact_ent}

We now discuss the dynamical behavior of entanglement, under closed as well as open system dynamics, where in the latter case, the initial state of the system is prepared to be a separable one, obtained by choosing parameters from the FS with a very low system temperature.

\subsection{Closed evolution}
\label{subsec:dyn_closed}

Similar to the CES, the density matrix corresponding to an even-odd NN spin-pair of the time-evolved state of the ATXY model with arbitrary $N$, in the case of closed system evolution, can be obtained analytically  using the single-site magnetizations and two-site spin correlation functions. However, unlike the CES, $T^{xy}_{eo}$ and $T^{yx}_{eo}$ do not vanish in the present case, and the density matrix corresponding to the NN even-odd spin pair is given by  \cite{chanda16}
\begin{eqnarray}
\hat{\rho}^{eo}(t) &=& \frac{1}{4}\Big[\mathbb{I}_e\otimes\mathbb{I}_o
 +m^z_e (\hat{\sigma}^z_e\otimes\mathbb{I}_o) + m^z_o  (\mathbb{I}_e\otimes  \hat{\sigma}^z_o)\nonumber\\
 &&+\sum_{\alpha=x,y,z}T^{\alpha\alpha}_{eo}(\hat{\sigma}^\alpha_e\otimes\hat{\sigma}^\alpha_o)  + T^{xy}_{eo}(\hat{\sigma}^x_e\otimes\hat{\sigma}^y_o)  \nonumber \\
 &&+ T^{yx}_{eo}(\hat{\sigma}^y_e\otimes\hat{\sigma}^x_o) \Big],
\label{rhoij}
\end{eqnarray}
where $m^z_{e(o)} = \mbox{Tr}(\hat{\sigma}^z_{e(o)} \hat{\rho}(t))$, $T^{\alpha\beta}_{eo} = \mbox{Tr}(\hat{\sigma}^{\alpha}_e \otimes \hat{\sigma}^{\beta}_o \hat{\rho}(t))$,  $\alpha, \beta = x, y, z$ can be computed analytically using the fermionic operators \cite{chanda16}. In our calculations, the initial state is chosen to be the CES with high $\beta_S$ and with other parameters satisfying Eq. (\ref{eq:fact_line}), having vanishing entanglement.

\begin{figure}
\includegraphics[scale=0.6]{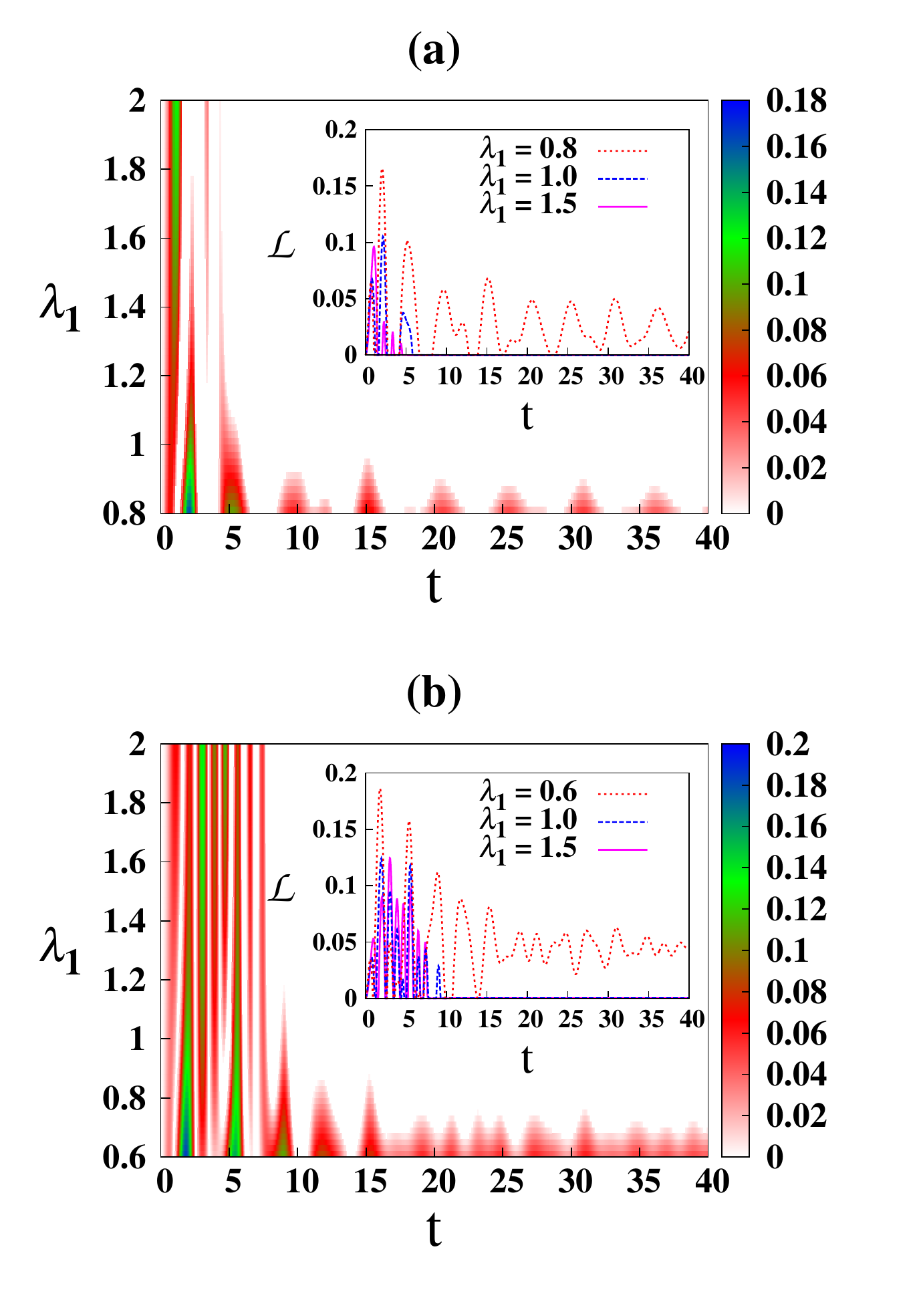}
\caption{(Color online.) \textit{Propagation of thermal entanglement after starting off from the factorization line under closed unitary evolution.} The variation of LN as a function of $t$ and $\lambda_1$ with (a) $\gamma=0.6$, and (b) $\gamma=0.8$, where $\lambda_2$ is fixed by Eq. (\ref{eq:fact_line}). (Insets) Variation of LN as a function of $t$ for different values of $\lambda_1$. The axes in all the figures are dimensionless. }
\label{fig:thermal_cd}
\end{figure}

With initial states that are not factorized, it was shown that NN entanglement under time-dependent magnetic field as given in Eq. (\ref{eq:magnetic_field}) oscillates and saturates to a positive value \cite{chanda16}. However, this is not the case if the dynamics starts from the separable state. Specifically, for $t > 0$, in the NN spin-pair, entanglement is created for high values of $\gamma$,  irrespective of $\lambda_1$. It then oscillates between zero and non-zero values during the initial phase of the dynamics. However, the oscillation quickly dies out and the LN vanishes for relatively high values of $\lambda_1$, while for lower values of $\lambda_1$, the oscillation sustains longer, and the value of LN even saturates to a non-zero value at large $t$. Such an analysis on $(\lambda_1, \lambda_2, \gamma)$-space reveals that LN, surviving for a large time, can only be obtained when the model is close to the UXY model, i.e., $\lambda_2 = 0$, $\lambda_1 \neq 0$, $\gamma > 0$. It is also visible from the insets of Figs. \ref{fig:thermal_cd}(a)-(b), where the variations of LN are plotted as a function of $t$ only, for different values of $\lambda_1$ and a fixed value of $\gamma$. Also, for higher values of $\gamma$, initial oscillation of entanglement for higher values of $\lambda_1$ sustains longer, as depicted in Figs. \ref{fig:thermal_cd}(a)-(b).  

\begin{figure}
\includegraphics[scale=0.335]{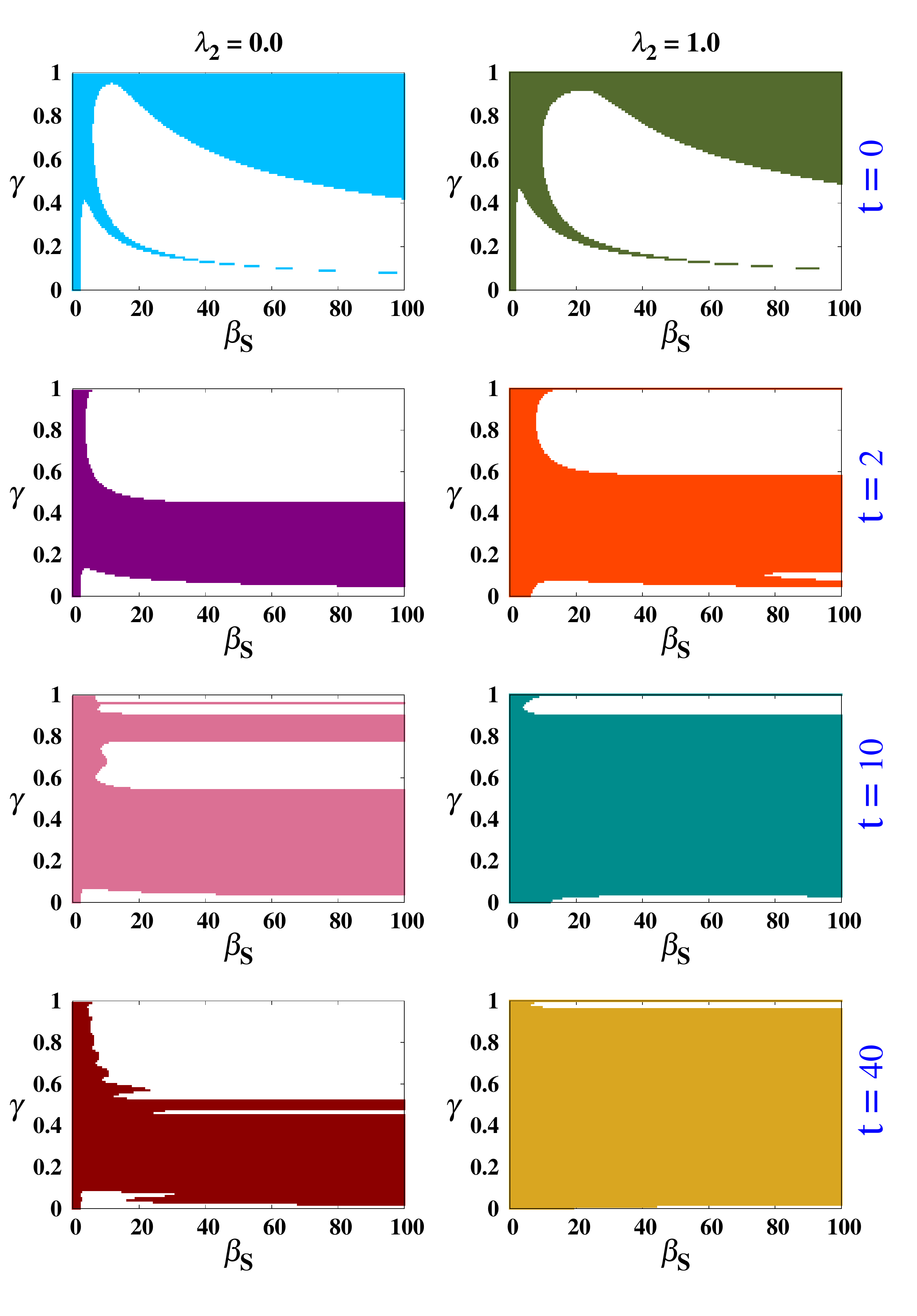}
\caption{(Color online.) \textit{Frozen-time snapshots of the $\mathcal{L}\neq0$ regions on the $(\beta_S,\gamma)$-plane.} The shaded regions in the figures represent the regions on the $(\beta_S,\gamma)$-plane where $\mathcal{L}=0$ while the white regions represent $\mathcal{L} \neq 0$. The left column of figures correspond to the UXY model ($\lambda_2=0$), while the right column is for the ATXY model ($\lambda_2=1$). The snapshots are taken at $t=0,2,10$ and $40$. The value of $\lambda_1$ is fixed by Eq. (\ref{eq:fact_line}) for all the points on the $(\beta_S,\gamma)$-plane. All quantities plotted are dimensionless.}
\label{fig:thermal_bg}
\end{figure} 
We now investigate how the landscape of thermally emergent entanglement on the $(\beta_S,\gamma)$-plane evolves with time under closed evolution. In order to do so, in Fig. \ref{fig:thermal_bg}, we map the regions of $\mathcal{L}\neq 0$ (white region) on the $(\beta_S,\gamma)$-plane at different instances of time, where the values of $\lambda_2$ are fixed, and the values of $\lambda_1$ are determined from Eq. (\ref{eq:fact_line}). The double-humped entanglement-pattern for $\gamma \leq 0.45$, as discussed in Sec. \ref{sec:ent_fact_line}, sustains only during the short-time dynamics. With increasing $t$, this feature disappears rather quickly (during $t\leq2$), while  regions of $\mathcal{L}\neq0$ may emerge on the $(\beta_S,\gamma)$-plane (for example, $t=2,10$) at specific time instances. Moreover, Fig. \ref{fig:thermal_bg} reveals a clear distinction between the UXY and ATXY model provided the initial state is chosen from the factorization-surface. Specifically, we observe that for sufficiently high $t$ (such as $t=10,40$), there exists substantial regions with $\mathcal{L}\neq 0$ on the $(\beta_S,\gamma)$-plane for the UXY model, while in case of the ATXY model, such $\mathcal{L} \neq 0$ region almost does not exist, i.e., $\mathcal{L}$ vanishes almost everywhere, except small regions at very high ($\geq0.95$), or very low $(\leq0.02)$ values of $\gamma$ and low value of the initial system temperature $\beta_S$.

\begin{figure}
\includegraphics[scale=0.345]{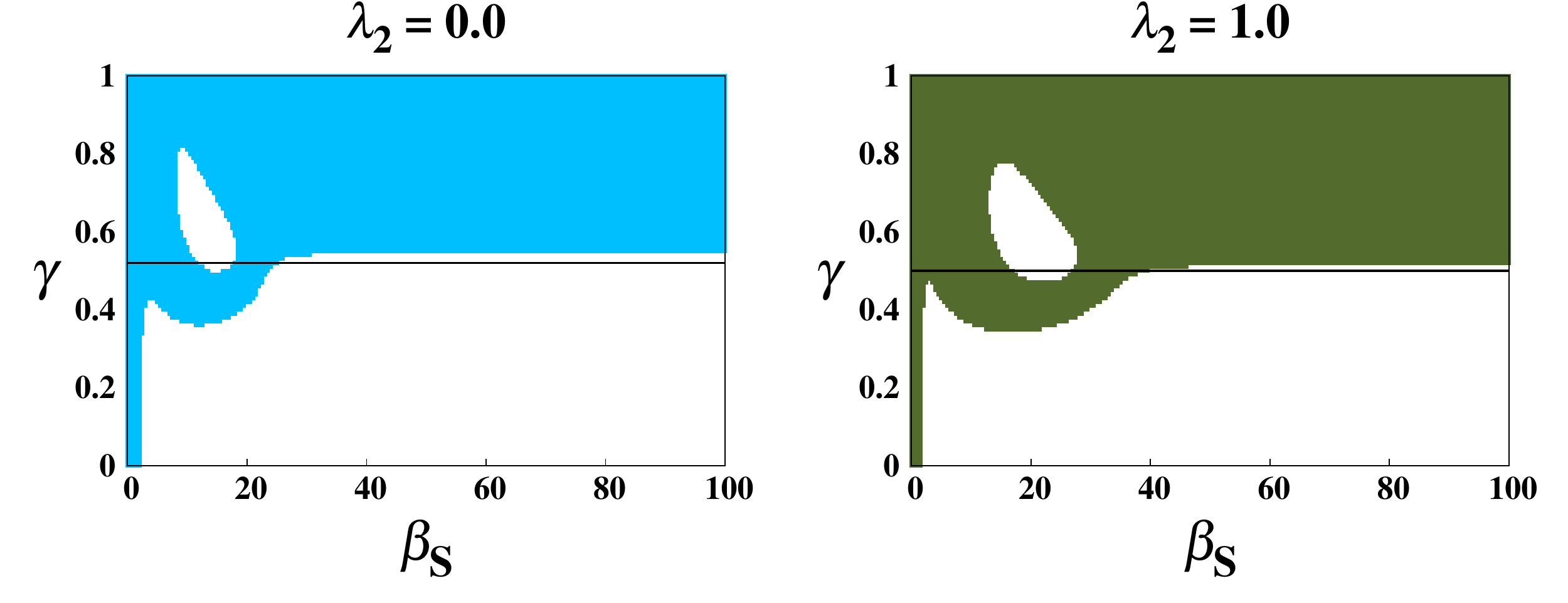}
\caption{(Color online.) \textit{Snapshots of the $\mathcal{L}\neq0$ regions on the $(\beta_S,\gamma)$-plane at $t=0$ for finite size system, specifically for $N = 10$.} The shaded regions in the figures represent the regions on the $(\beta_S,\gamma)$-plane where $\mathcal{L}=0$. The left figure corresponds to the UXY model ($\lambda_2=0$), while the right one is for the ATXY model ($\lambda_2=1$). The value of $\lambda_1$ is fixed by Eq. (\ref{eq:fact_line})
on the $(\beta_S,\gamma)$-plane. The horizontal lines in the figures represent the model with $\gamma = 0.5$, 
 where a double revival of LN takes place with varying $\beta_S$ (compare with Fig. \ref{fig:thermal_static}(a)), mimicking the behavior of entanglement of the model in the thermodynamic limit. All quantities plotted are dimensionless.}
\label{fig:thermal_bgN10}
\end{figure} 

We point out here that all the results discussed above correspond to the system described by the Hamiltonian $\hat{H}_S$ in the thermodynamic limit. However, in the succeeding section, when we consider the system to be exposed to an environment, we can only address this question for finite system size. Before proceeding towards this, it is important to consider how the features of the closed dynamics changes, when the system consists of finite number of spins, $N$. In the finite-sized system, FS remains unchanged, while the phase-boundaries change only slightly. Since the change is small-enough, Eqs. (\ref{eq:pmafm}) and (\ref{eq:dmafm}) can be considered as the effective phase-boundaries in the finite-size scenario. However, the double revival of entanglement with varying $\beta_S$ over the FL at $t=0$ is absent for small system sizes, and there is a single $\mathcal{L}\neq0$ region on the $(\beta_S,\gamma)$-plane. However, for $N\geq10$, a second region of non-zero LN at lower values of $\beta_S$, and consequently the double revival appears. The region of $\mathcal{L}\neq0$ at low values of $\beta_S$ starts growing with the increase of the system size. An example of double-revival in the case of $N=10$ is depicted in Fig. \ref{fig:thermal_bgN10}. However, we observe that at large-time ($t\geq10$), the regions of non-vanishing entanglement on the $(\beta_S,\gamma)$-plane, and the oscillatory behavior of LN on the $(t,\lambda_1)$-plane for different values of $\gamma$ qualitatively match with those in the case of $N\rightarrow\infty$.

\subsection{Open system dynamics}
\label{subsec:dyn_open}

We now focus on the dynamics of the quantum spin model, described by the Hamiltonian $\hat{H}_S$, in contact with a thermal bath acting as an environment to the system. As the bath, we consider a collection of identical and decoupled spins \textcolor{red}{\cite{rqim_attal, rqim}}, each at a inverse temperature $\beta_E=1/(k_BT_E)$ and described by the Hamiltonian $\hat{H}_E=B\hat{\sigma}^z_E$, with $B$ being the energy of one qubit. The interaction of the reservoir with the system is such that during a very small time interval $\delta t$, only one spin from the collection interacts with a ``chosen" spin in the system, labeled as the ``door", via the interaction Hamiltonian given by 
\begin{eqnarray}
\hat{H}_{int}=k^{1/2}\delta t^{-1/2}(\hat{\sigma}^x_{d}\hat{\sigma}^x_E+\hat{\sigma}^y_{d}\hat{\sigma}^y_E),
\end{eqnarray}
where $k$ has the dimension of (energy${}^2 \times$ time) , and the subscript ``d" indicates the door spin. 
In each such small time intervals of duration $\delta t$, one spin from the collection interacts with one spin from a system via 
 the door, thereby giving rise to a \textit{repetitive} interaction between the bath and the system \textcolor{red}{\cite{rqim_attal, rqim}}. In a more general ``multidoor" scenario, a number of independent environments may interact with a number of chosen spins in the system. In such a case, the interaction Hamiltonian is of the form $\hat{H}_{int}=k^{1/2}\delta t^{-1/2}\sum_{l=1}^{N_d}(\hat{\sigma}^x_{d_{l}}\hat{\sigma}^x_E+\hat{\sigma}^y_{d_{l}}\hat{\sigma}^y_E)$, where $N_d$ is the number of doors. The quantum master equation that dictates the dynamics of the system for single door is given by 
\begin{eqnarray}
\dot{\hat{\rho}}_S=-\frac{i}{\hbar}[\hat{H}_S,\hat{\rho}_S]+\mathcal{D}(\hat{\rho}_S), 
\label{eq:qme}
\end{eqnarray}
where 
\small
\begin{eqnarray}
\mathcal{D}(\hat{\rho}_S)&=& \frac{2k}{{\hbar}^2 Z_E}\sum_{l=1}^{N_d}\sum_{i=0}^1 e^{(-1)^i \beta_E B}[2\hat{\eta}_{d_l}^{i+1}\hat{\rho}_S\hat{\eta}_{d_l}^i-\{\hat{\eta}_{d_l}^i\hat{\eta}_{d_l}^{i+1},\hat{\rho}_S\}], \nonumber \\
\label{eq:dynamical_term}
\end{eqnarray}\normalsize
with $Z_E=\mbox{Tr}[\exp -\beta_E \hat{H}_E]$, and $\hat{\eta}_{d_l}^\alpha=\hat{\sigma}_{d_l}^x+(-1)^\alpha\hat{\sigma}_{d_l}^y$ \cite{rqim, chanda16a}. Another dimensional analysis suggests that for the Hamiltonian $\hat{H}_S$ and with $D(.)$ given in Eq. (\ref{eq:dynamical_term}), time $t$ in Eq. (\ref{eq:qme})  is in the unit of $\hbar /J$, and $k$ is in the unit of $\hbar J$. We therefore redefine the dimensionless quantities $k$ and $t$ as  $k\rightarrow k/\hbar J$ and $t\rightarrow tJ/\hbar$ respectively, and use them throughout the paper. For the purpose of our calculation, we set the dimensionless quantity $k=1$. Note here that the $i=0$ terms in Eq. (\ref{eq:dynamical_term}) represent the dissipation process with rate $Z_E^{-1}\exp(\beta_E B)$, while the terms with $i=1$ are for absorption process with rate $Z_E^{-1}\exp(-\beta_E B)$. In the case of high values of $\beta_E B (\beta_E B\geq5)$, the rate of the absorption process becomes negligible, and the dynamical term in Eq. (\ref{eq:dynamical_term}) represents that of an amplitude-damping channel under Markovian approximation \cite{petruccione}. Unless otherwise stated, we keep $\beta_E B=10$ for all our calculations throughout this paper, and hence neglected the $i=1$ term.

We determine $\hat{\rho}_S$ as a function of $t$ by numerically solving Eq. (\ref{eq:qme}) for specific values of $N$, and then trace out all the spins except a NN even-odd pair to obtain the reduced state corresponding to the chosen pair.  This reduced state can, in turn, be used to compute the NN LN as a function of $t$. We assume that the system is initially prepared in a thermal equilibrium state, $\hat{\rho}_{eq}(t=0)$, with a heat bath at a very low temperature at $t=0$, at which point the repetitive quantum interaction is turned on. 
Evidently, the initial state, and thereby the dynamics depends on the choice of the parameters of $H_S$ at $t=0$, given by $\{\gamma,\lambda_1,\lambda_2, \beta_S\}$. Choice of the values of system parameters from different phases of the model gives rise to a rich variety of dynamics.

\begin{figure}
\includegraphics[scale=0.35]{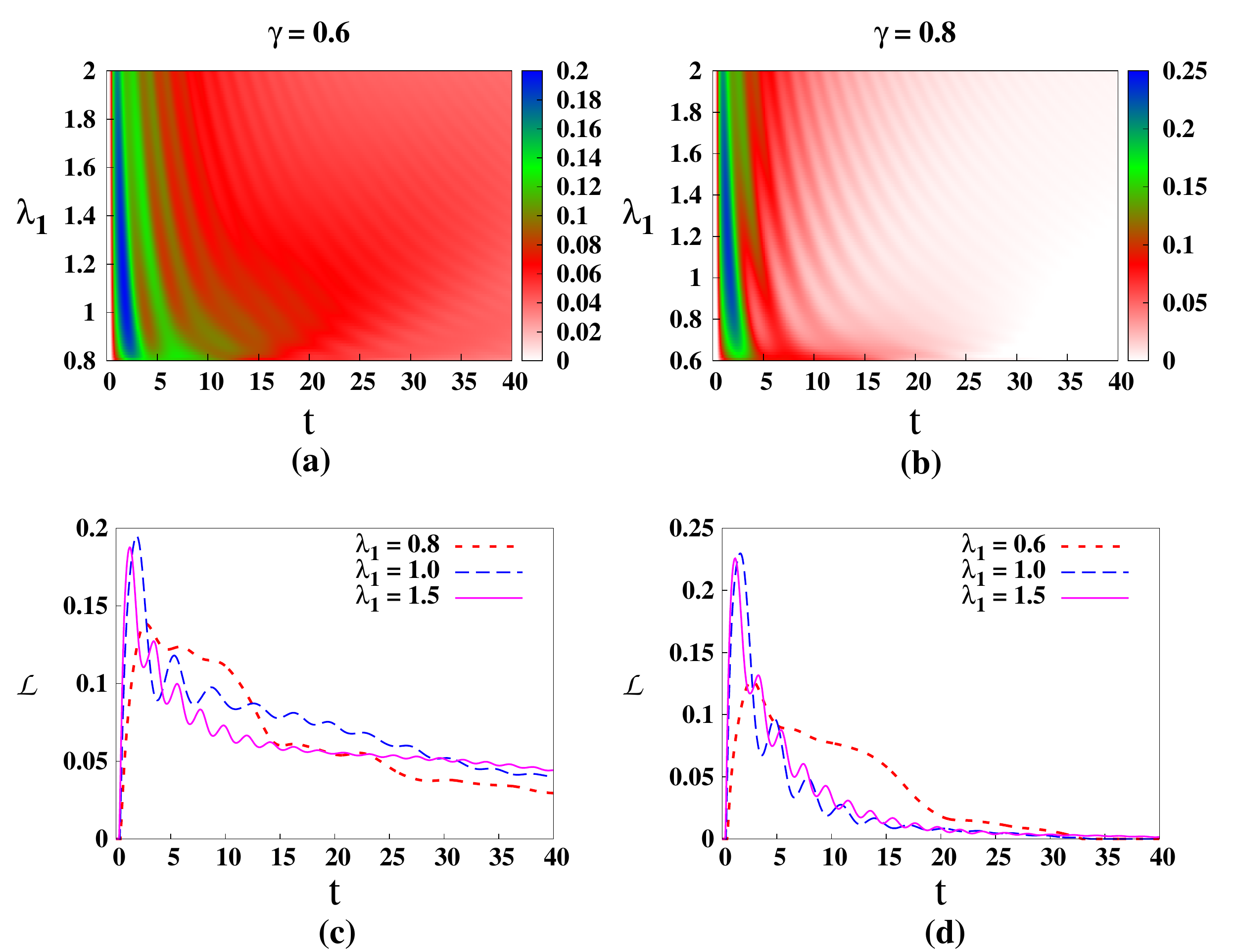}
\caption{(Color online.) \textit{Open system dynamics of entanglement under repetitive quantum interaction after starting off from the factorization line.} The variation of LN as a function of $t$ and $\lambda_1$ with (a) $\gamma=0.6$, and (b) $\gamma=0.8$. The variations of LN as a function of $t$ for different values of $\lambda_1$ are given in (c) for $\gamma=0.6$ and (d) for $\gamma=0.8$. Entanglement generation under closed vs. open dynamics can be made by comparing insets of Figs. \ref{fig:thermal_cd} (a)-(b) and (c)-(d) in above figures. Although in a closed unitary evolution, entanglement can be preserved for a long time while it is not possible in an open dynamics considered in this paper.
All quantities plotted are dimensionless.}
\label{fig:thermal_open}
\end{figure} 

We demonstrate the results considering the single-door scenario $(N_d=1)$ and a spin-chain of size $N$ under periodic boundary condition. Without loss of generality, let us label the spins of the system as $\{1, 2, \ldots,N\}$, where we assume that the first spin interacts with the bath via door. For ease of discussion, let us divide set of spins in the system into two mutually disjoint sets. The first set, $\mathcal{S}_1$, consists of all the NN spin-pairs each of which contains at least one door spin, while the second set, $\mathcal{S}_2$, is constituted of all the NN spin-pairs none of which contains a door spin. Clearly, $\mathcal{S}_1$ consists of two NN spin-pairs, i.e., $\mathcal{S}_1\equiv\{(1,2),(N,1)\}$, while $\mathcal{S}_2$ is constituted of the  rest of the NN spin-pairs, $\mathcal{S}_2\equiv\{(i, i+1);\;2\leq i\leq N-1\}$.   We begin our discussion with the latter set, and take the NN spin-pair, say, $(2, 3)$ as an example in the case of a spin-chain with $N=10$. In the same spirit as in the case of the closed dynamics, we choose the values of the system parameters according to the FS. The environment temperature, $\beta_E(=10)$ is moderately high compared to the value of $\beta_S$, set at $\beta_S=80$, which can faithfully mimic the low-temperature ($\beta_S\rightarrow\infty$) properties of the model at $N=10$. Interestingly, for a fixed value of $\gamma$, LN is found to be generated over a very small region on the $(t,\lambda_1)$-plane ($0.75\leq \lambda_1\leq 0.9;\; 0\leq t \leq 10$), while the values of $\lambda_2$ are fixed by Eq. (\ref{eq:fact_line}). Also, the value of the NN LN generated over the spin-pair $(2,3)$ is $\mathcal{L}\leq 8\times 10^{-2}$. This suggests that the amount and duration of entanglement generation is very small for the spin-pairs belonging to $\mathcal{S}_2$ if the system parameter values corresponding to the initial state of the open system dynamics is chosen according to Eq. (\ref{eq:fact_line}). Note here that the FL is encompassed completely in the AFM phase of the model. 

The situation becomes drastically different in the case of $\mathcal{S}_1$. Figs. \ref{fig:thermal_open} (a)-(b) depict the variation of the LN for the spin-pair $(1, 2)$, which is same as $(N, 1)$ due to periodicity, as a function of time and $\lambda_1$ with (a) $\gamma=0.6$ and (b) $\gamma=0.8$.  The values of $\lambda_2$ are fixed by the factorization condition, and the values of $\beta_S$ and $\beta_E$ are the same as those used in the former case. It is clear from the figures that considerable entanglement is generated during the dynamics, with the maximum value of $\mathcal{L}$ increasing with increasing $\gamma$.  LN corresponding to the spin-pair $(1,2)$ sustains for a longer time compared to the former case of $\mathcal{S}_2$. The duration in which $\mathcal{L} \neq 0$  decreases with increasing $\gamma$, as can be seen from the figures, indicating a trade-off between the generation of higher values of entanglement and the length of the time interval in which $\mathcal{L}\neq0$. A clearer picture can be obtained from Figs. \ref{fig:thermal_open}(c)-(d), where the variation of LN as a function of time, corresponding to two specific values of $\lambda_1$ for each values of $\gamma$ is shown. Also, note that with a fixed value of $\gamma$, entanglement oscillates at first, and then decays to zero irrespective of the values of $\lambda_1$. This behavior is in contrast with the same in the case of closed dynamics, where entanglement is found to saturate at a non-zero value for lower values of $\lambda_1$. Moreover, we observe that with the increase of $N$, the decay rate of entanglement becomes slower although the qualitative behavior of entanglement with time remains unaltered.

We point out here that by using CES with non-zero entanglement corresponding to the system parameter values not belonging to the FL, and chosen from the PM-I, PM-II, and AFM phases as initial states, NN LN can remain invariant with time for a finite duration -- a phenomenon known as the \textit{freezing} of entanglement \cite{chanda16a}. Interestingly, freezing of entanglement is observed only in the NN spin-pairs belonging to $\mathcal{S}_2$, while the dynamics of NN LN corresponding to the spin-pairs belonging to $\mathcal{S}_1$ is highly oscillatory. Note here that similar to the freezing of entanglement, generation of entanglement during open system dynamics, where the system parameters are chosen from the FS, clearly distinguishes between the two sets of spin-pairs, $\mathcal{S}_1$ and $\mathcal{S}_2$. However, in contrast to the freezing of entanglement, the spin-pairs belonging to $\mathcal{S}_1$ provides a more beneficial situation in terms of emergence of NN entanglement over initially unentangled states by the action of environmental noise, as discussed above.

All of the results regarding open dynamics of the system discussed so far correspond to a high value of $\beta_S$ $(=80)$, and a relatively low value of $\beta_E B (=10)$. We conclude the discussion on open system dynamics by pointing out that for fixed $\beta_E B=10$, the qualitative features of all the above  results remain unchanged  even with a varying $\beta_S$ except when the system temperature is high ($\beta_S\leq 10$). In that case, almost no entanglement is generated throughout the dynamics, irrespective of the sets $\mathcal{S}_1$ and $\mathcal{S}_2$, when the initial state is factorized. Also, for fixed $\beta_S=80$, one can explore lower values of $\beta_E B$, where the absorption terms in the quantum master equation becomes non-vanishing. However, the qualitative features of the dynamics of NN LN corresponding to the spin-pairs belonging to the sets $\mathcal{S}_1$ and $\mathcal{S}_2$ remains unchanged. Moreover, similar observations are found when the system-environment interaction is considered in the multidoor scenario. 

We conclude by mentioning that the noise model used in the above discussions is a local one of dissipative type. However, one can also consider a non-dissipative noise, such as the local dephasing, instead of a dissipative one using the same formalism. We find that generation of entanglement during the open system dynamics of the model, with the initial state corresponding to the system parameters satisfying FS, is possible for non-dissipative noises like the dephasing noise also.

\section{Conclusion}
\label{sec:conclud}

In certain quantum many-body systems, system parameters, chosen in a specific way, leads to a zero-temperature state that is product across any bipartition, known as a factorized state. In the entanglement resource theory, where entanglement is used as resource for different quantum informations processing schemes, such states are useless. At the same time, spin models are turned out to be appropriate physical systems for realizing quantum information protocols which can be realized in the laboratory. One possibility to avoid such factorized states is to create the system  far from the factorized region. If such control over the system-preparation is missing, we can ask whether  entanglement can be generated by tuning the system temperature, or by considering the closed and open dynamics of the system, in quantum states that correspond to the factorization points. It is important to note at this point that reaching absolute zero temperature is hard compared to the preparation of a system with moderate temperature.  Also, evolution of a system with time, under closed setup, or in contact with an environment, can be a natural choice for quantum information processing.

For such investigation, we choose an one-dimensional anisotropic quantum XY model in the presence of a uniform and an alternating transverse magnetic field. For fixed values of the anisotropy parameter, the factorization points of this model are known to form two lines \cite{chanda16} on the plane of  relative strengths of the uniform and transverse magnetic fields and the zero-temperature states are unentangled over these lines. We show that by increasing the temperature of the system in canonical equilibrium state, double revival of entanglement happens when value of the anisotropy parameter is chosen in an appropriate range. Although the non-monotonic behavior of entanglement with the equilibrium temperature in quantum spin-models, and the single revival of thermal entanglement with increasing temperature were known \cite{chanda16,arnesen}, the existence of a double revival of thermal entanglement is counter-intuitive, and has not been reported earlier. Interestingly, such double-humped behavior of entanglement occurs when one starts from the thermal state corresponding to the factorization line.

We also show that under closed unitary evolution of the system driven out of equilibrium by a sudden change in the system parameters, namely, the magnetic fields, considerable entanglement is generated during the dynamics. The initial state is separable, prepared by choosing system parameters from the FS. The results indicate that a low value of uniform magnetic field in the ATXY model is favorable for sustaining generated entanglement in the long time limit, while the entanglement oscillates and dies out rapidly for high value of the uniform magnetic field. On the other hand, when the system interacts with an external thermal bath via a repetitive quantum interaction, entanglement of certain nearest-neighbor spin-pair persists for all values of the uniform field when the value of the anisotropy parameter is low, but dies out quickly when the anisotropy is increased.  The open system dynamics also distinguishes between the spin-pairs that have a direct connection with the external bath and the spin-pairs that have not. Counterintuitively, entanglement in the spin-pair which is in contact with a thermal bath has high value and long duration compared to the spin-pairs which do not interact with the bath. Moreover, we find that duration of non-vanishing entanglement and the amount of entanglement in this scenario has complementary relation. Such generation of entanglement is also possible for other environments like the ones resulting in local dephasing noise etc. Apart from the entanglement creation, such study  reveals the variation of entanglement due to the interplay between system parameters, temperatures, environments.


\begin{thebibliography}{100}
\bibitem{dimertoPMII} Note that in the earlier papers \cite{chanda16, dutta-qpt-books, alt-field-old}, the PM-II phase was mentioned as the dimer phase. Our recent analysis \cite{amader_dmpaper} shows that the dimer order parameter \cite{dm_order} in this phase vanishes and it is indeed paramagnetic in nature. However, such finding does not affect the results obtained in this paper as well as the other results in \cite{chanda16, dutta-qpt-books, alt-field-old}.
\end{thebibliography}
\end{document}